\documentclass[10pt]{iopart}
\usepackage{graphicx}
\usepackage{color}
\usepackage{balance}
\usepackage{cite}
\usepackage{hyperref}

\hypersetup{
    colorlinks=true,
    linkcolor=black,
    citecolor=black,
    filecolor=black,
    urlcolor=black,
}

\begin{document}

\title[A tunable shunt-enabled rainbow trap]{Manipulating waves by distilling frequencies:\\a tunable shunt-enabled rainbow trap}

\author{Davide Cardella$^{*}$, Paolo Celli$^{*}$, Stefano Gonella}

\address{Department of Civil, Environmental, and Geo- Engineering, University of Minnesota, Minneapolis, MN, 55455, USA}
\vspace{5pt}
\address{$^{*}$ D.C. and P.C. both contributed majorly to the realization of this work}
\ead{sgonella@umn.edu}
\vspace{10pt}
\begin{indented}
\item[]{\underline{\bf Published article}: \emph{Smart Materials and Structures} {\bf 25} (8), 085017 (2016);\quad \url{http://dx.doi.org/10.1088/0964-1726/25/8/085017}}
\end{indented}

\begin{abstract}
In this work, we propose and test a strategy for tunable, broadband wave attenuation in electromechanical waveguides with shunted piezoelectric inclusions. Our strategy is built upon the vast pre-existing literature on vibration attenuation and bandgap generation in structures featuring periodic arrays of piezo patches, but distinguishes itself for several key features. First, we demystify the idea that periodicity is a requirement for wave attenuation and bandgap formation. We further embrace the idea of ``organized disorder'' by tuning the circuits as to resonate at distinct neighboring frequencies. In doing so, we create a tunable ``rainbow trap''~[Tsakmakidis \emph{et al} 2007 \emph{Nature} {\bf 450} 397–401] capable of attenuating waves with broadband characteristics, by \emph{distilling} (sequentially) seven frequencies from a traveling wavepacket. Finally, we devote considerable attention to the implications in terms of packet distortion of the spectral manipulation introduced by shunting. This work is also meant to serve as a didactic tool for those approaching the field of shunted piezoelectrics, and attempts to provide a different perspective, with abundant details, on how to successfully design an experimental setup involving resistive-inductive shunts.
\end{abstract}

%
\vspace{2pc}
\noindent{\it Keywords}: Wave manipulation, Resonant shunts, Metamaterials, Rainbow trap, Broadband filter
%
%
%
\ioptwocol

\section{Introduction}
\label{sec:intro}

\begin{figure*} [!t]
\centering
\includegraphics[scale=1.38]{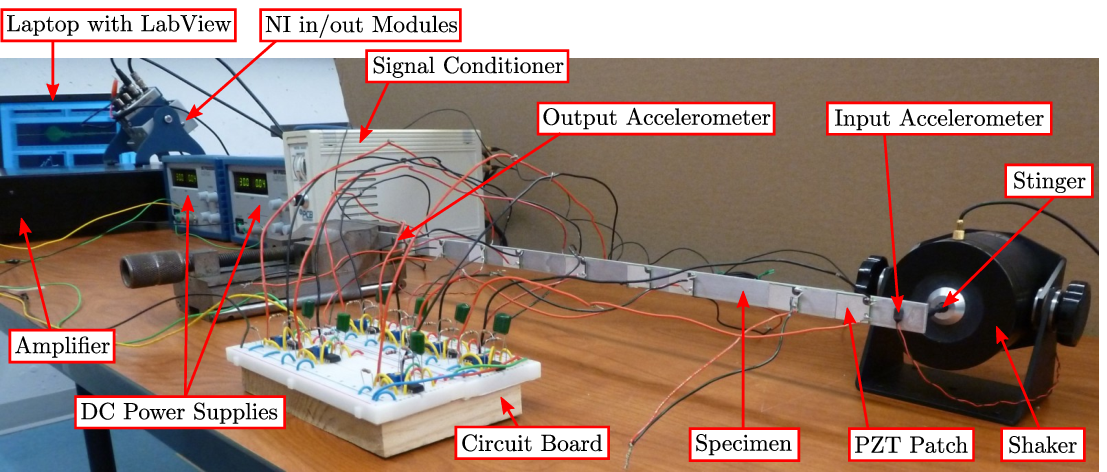}
\caption{Experimental setup.}
\label{fig:setup}
\end{figure*}
\begin{figure*} [!htb]
\centering
\includegraphics[scale=1.38]{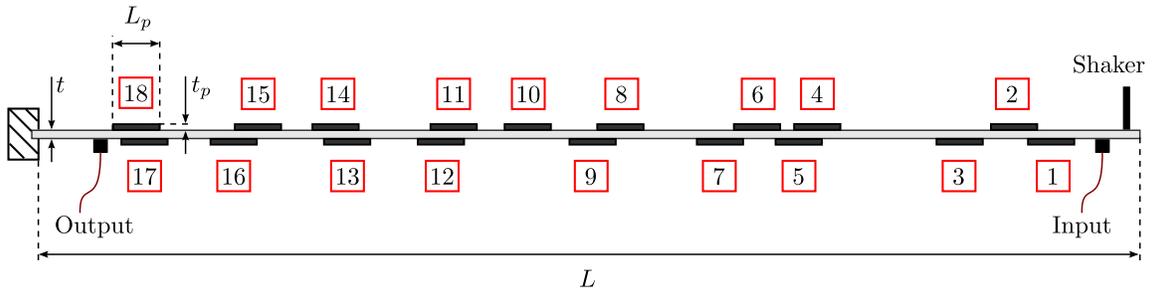}
\caption{Schematic (side view) of the beam specimen, highlighting some of its characteristic dimensions. This figure is to be used as reference for the numbering of the patches. The solid black squares represent input and output accelerometers.}
\label{fig:beam}
\end{figure*}

Ever since Robert Forward's paper on the electronic damping of vibrations in optical structures~\cite{Forward_APPLOPT_1979} (where he stated ``We hope that the examples presented in this paper will [...] lead to the realization that one does not need to find a purely mechanical solution to a mechanical vibration problem''), there has been extensive progress in the field of vibration control via shunted piezoelectric elements. \emph{Shunting} is referred to as the act of connecting the electrodes of a piezoelectric element to a passive electrical circuit, which represents an electrical extension of the structure. A seminal work on the topic is that of Hagood and von Flotow~\cite{Hagood_JSV_1991}, who presented a detailed theoretical formulation on shunting based on linear piezoelectric theory, and experimental results on the attenuation of a single mode of vibration via resistive and resistive-inductive (RL) circuits. While resistive circuits behave as broadband dampers, RL circuits allow to obtain pronounced and localized vibration attenuation; when connecting a RL in series to a piezo patch (which can be modeled as a capacitor in series with a voltage supply) we obtain a resonant RLC circuit, which behaves as an electrical dynamic damper which can be tuned by modifying the circuit parameters. In recent years, extensive efforts have been devoted to extending the effects of RL circuits to multiple modes of vibration~\cite{Hollkamp_JIMSS_1994, Viana_JBSMSE_2006, Thomas_SMS_2012} (multimodal vibration attenuation) and to the investigation of alternative passive circuits, such as the negative capacitor (NC)~\cite{Behrens_SMS_2003, Park_JVC_2005, Beck_JIMSS_2011}. Moreover, with the discovery of phononic crystals~\cite{Kushwaha_PRL_1993}---periodic structures with unconventional wave manipulation characteristics such as phononic bandgaps---researchers have begun to incorporate shunted piezoelectric patches in the architectures of these materials to obtain tunable, complex wave patterns. This has motivated an increasingly larger number of studies on structures involving \emph{periodic arrays} of shunted piezos. Among these, we recall the pioneering contribution of Thorp et al.~\cite{Thorp_SMS_2001}, who paired the concepts of phononic crystals and shunting for the first time. It is also interesting how the authors hinted at the concept of \emph{relaxation} of the periodicity to widen the bandgaps, invoking \emph{Anderson localization} mechanisms, anticipating a discussion that has become central in recent phononic crystals literature~\cite{Celli_APL2_2015}. Noteworthy is also the work of Airoldi and Ruzzene~\cite{Airoldi_NJoP_2011}, who were the first to use a periodic structure with independent RL shunts to control waves, rather than to attenuate resonance peaks; their work---the first catering to problems encountered in metamaterial engineering applications---provided compelling results in terms of locally-resonant bandgap generation. In parallel, researchers have also proposed more convoluted active and passive shunting strategies to enhance vibration attenuation. Examples are electrical networks placed in parallel to the mechanical network (where each patch is shunted to the neighboring patch instead of connected to ground)~\cite{dellIsola_SMS_2004, Lossouarn_SMS_2015, Bergamini_JAP_2015}, enhanced resonant shunts~\cite{Wang_SMS2_2011} and amplifier-resonator feedback circuits~\cite{Wang_SMS_2016}. While most of the studies involving periodic arrays of shunted piezos are centered around vibration attenuation and bandgap generation~\cite{Thorp_SMS_2001, Airoldi_NJoP_2011, dellIsola_SMS_2004, Spadoni_JIMSS_2009, Casadei_SMS_2010, Airoldi_JIMSS_2011, Wang_SMS_2011, Wang_SMS2_2011, Bergamini_ADVMAT_2014, Chen_JVA_2014, Lossouarn_SMS_2015, Bergamini_JAP_2015, Zhou_SMS_2015, Wang_SMS_2016, Zhu_APL_2016}, others have proposed shunted piezos as a mean to achieve tunable wave focusing in metamaterial architectures~\cite{Celli_APL_2015, Wen_JIMSS_2016, YI_SMS_2016}, and tunable waveguiding~\cite{Oh_APL_2011, Casadei_JAP_2012}.

So far, most of the experimental studies involving multiple patches bonded to the same substrate have been characterized by a uniform tuning of the circuit characteristics. Apart from the previously-discussed work by Thorp et al.~\cite{Thorp_SMS_2001}, exceptions are the recent works of Wang and Chen~\cite{Wang_SMS_2016} and Bergamini et al.\cite{Bergamini_JAP_2015}, who discussed the use of a superlattice method and diatomic unit cells, respectively, as strategies to extend the bandgaps to multiple frequencies. In both cases, different inductance values are connected to neighboring patches, while preserving some sort of periodicity. 
While modeling structures with periodically-placed piezoelectric elements might present some convenient aspects, such as the possibility of using reduced models (i.e., a unit cell analysis) and the availability of additional frequency regions of Bragg attenuation, one might argue that, when dealing with resonant shunts, periodicity is not strictly needed for bandgap generation and wave attenuation. In fact, when locally-resonant mechanisms are involved, the resonators' placement throughout the medium does not affect the global formation of bandgaps~\cite{Rupin_PRL_2014, Celli_APL2_2015}. On the contrary, in some situations, it might be convenient to leverage some ``organized disorder'' to enhance the wave attenuation performance. The attempt to extend the usually narrow frequency regions of resonance-based attenuation has received growing interest in the metamaterials community over recent years. An example is the work of Kr\"odel et al.~\cite{Kroedel_EML_2015} on metamaterial-enabled seismic isolation, where a wide locally-resonant bandgap is created by employing resonators tuned at adjacent frequencies and displaying slightly overlapping spectra. The concept of trapping a wide spectrum of adjacent frequencies has been pioneered by work in optics, where the application to the control of visible light has inspired the name ``rainbow trap''~\cite{Tsakmakidis_NATURE_2007, Gan_PNAS_2011}, and has recently been extended to the realm of acoustic waves~\cite{Zhu_SCIREP_2013, Ni_SCIREP_2014, Zhou_APL_2016}.

In our work, we embrace these ideas of disorder and relaxation of the periodicity to demonstrate the wave attenuation capabilities of an array of randomly-positioned piezoelectric elements shunted with non-uniform RL circuits. We demonstrate that a device capable of attenuating vibrations over a broad frequency range can be obtained by tuning different RL circuits to resonate at different frequencies, thus realizing a tunable rainbow trap for elastic wave manipulation. The behavior of our test system is probed with a transient chirp signal and monitored by measuring how its spectro-spatial characteristics are modified by the sequential activation of the shunting circuits. Moreover, we propose an additional angle to the problem, whereby we look at the effects of wave manipulation through the prism of wave packet distortion.

This paper is organized as follows. In Sec.~\ref{sec:exp} we discuss in detail our experimental setup, including the specimen, the circuitry and the acquisition system. In Sec.~\ref{sec:res1} we discuss the individual tuning of the circuits. In Sec.~\ref{sec:res2} we report on the broadband wave attenuation enabled by the rainbow trap strategy. Finally, the conclusions of our work are drawn in Sec.~\ref{sec:con}.

\section{Experimental setup}
\label{sec:exp}

In this section, we describe in detail the experimental framework used throughout this work. A comprehensive picture of the setup is shown in Fig.~\ref{fig:setup}. We would like to point out that most of the circuitry solutions used here are inspired by the informative works of Viana and Steffen~\cite{Viana_JBSMSE_2006}, and of Thomas, Ducarne and De\"u~\cite{Thomas_SMS_2012, Ducarne_THESIS_2009}.

\subsection{Specimen geometry and patch configuration}
\label{sec:spec}
Our specimen is an Aluminum beam with eighteen PZT patches (STEMiNC, part SMPL20W15T14R111) bonded according to a random spatial pattern, as shown in the schematic of Fig.~\ref{fig:beam}. Note that the beam is instrumented with a dense population of patches. While this may be redundant for most wave control applications, it grants the possibility to test a variety of shunting sequences and to potentially investigate the effects of location, clustering and order. In our experiments, a maximum of seven patches will be activated simultaneously. Note that each patch has a slightly different value of capacitance; the average capacitance measured with a multimeter is $C_{p,\,ave}= 2.137\,\mathrm{nF}$.
The main geometric dimensions of the specimens are the following: the length of the beam is $L=51.6\,\mathrm{cm}$, the thickness of the beam is $t=0.26\,\mathrm{cm}$, the width of the beam is $b=1.5\,\mathrm{cm}$, the length of each patch is $L_p=2\,\mathrm{cm}$, the thickness of a patch is $t_p=0.14\,\mathrm{cm}$ and the width of the patch is $b_p=1.5\,\mathrm{cm}$. Note that, for the sake of brevity, we did not report the distance between each pair of neighboring patches; it will be demonstrated that this parameter is essentially uninfluential with respect to the effects discussed in this article. The patches are bonded to the beam using a 2-part epoxy glue (3M Scotch-Weld 1838 B/A); details on how to properly bond the patches are reported in the Supplementary Data (SD) section. The sketch also shows the position of the actuator and of the input and output accelerometers.

\begin{figure} [!htb]
\centering
\includegraphics[scale=1.38]{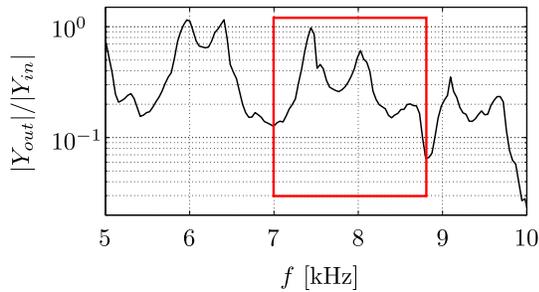}
\caption{Frequency response function of the beam specimen (with short-circuited patches), obtained using the signals recorded at the input and output accelerometers in response to a multi-tone signal in the $5$--$10\,\mathrm{kHz}$ range. The boxed region corresponds to the frequency interval of interest for our analysis.}
\label{fig:beamtf}
\end{figure}

The measured frequency response function of the beam (obtained using the signals recorded at the input and output accelerometers in response to a multitone excitation spanning the $5$--$10\,\mathrm{kHz}$ range) is shown in Fig.~\ref{fig:beamtf}.
The region highlighted in red represents the frequency interval of interest, i.e., the region where we want to test our wave attenuation strategy; note how this region, ranging from $7\,\mathrm{kHz}$ to $8.8\,\mathrm{kHz}$, deliberately spans multiple peaks of the beam response. The reason for selecting this frequency region will be explained in Sec.~\ref{sec:circ}.

\subsection{Circuitry}
\label{sec:circ}
\begin{figure} [!htb]
\centering
\includegraphics[scale=1.38]{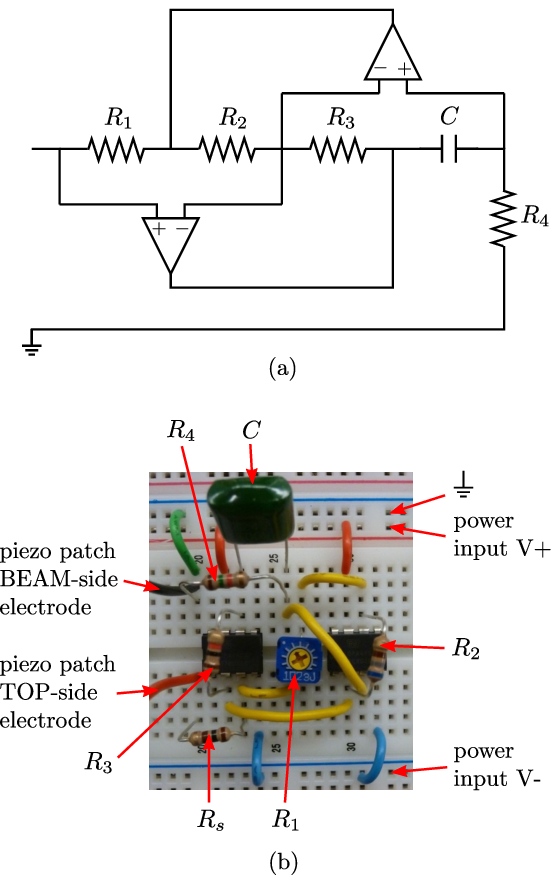}
\caption{The Antoniou circuit. (a) Schematic. (b) Realization on a breadboard, showing a series resistor $R_s$ and a tunable potentiometer in lieu of $R_1$.}
\label{fig:circ}
\end{figure}

Throughout this work, we resort to resistor-inductor (RL) shunting circuits. As shown by Hagood and von Flotow~\cite{Hagood_JSV_1991} and by Viana and Steffen~\cite{Viana_JBSMSE_2006}, a piezoelectric patch (which acts as a capacitor with capacitance $C_p$ in series with a voltage supply) placed in series with a RL circuit realizes a resonant RLC circuit and therefore behaves as a dynamic damper. In analogy with purely-mechanical resonant systems, the electric dynamic damper resonates at a frequency that can be determined from the circuit parameters as:
\begin{equation}
f_{res}=\frac{1}{2\,\pi}\sqrt{\frac{1}{L\,C_p}}\,\,\,\,\,.
\label{eq:res}
\end{equation}
Note that the resistance $R$ does not influence the natural frequency of the resonant circuit, but is directly proportional to the level of damping introduced in the structure. Given the very small capacitance values of the piezoelectric patches (of the order of $1\,\mathrm{nF}$), and given the fact that we want to operate at relatively low frequencies (of the order of $1000\,\mathrm{Hz}$), the inductances required to achieve resonant conditions in the desired range could vary in between $0.01\,\mathrm{H}$ and $10\,\mathrm{H}$. Such values of inductance are achievable with conventional coil-based inductors, using devices with characteristic sizes of the order of few centimeters. To avoid working with inductors of impractical dimensions, it is customary to resort to \emph{synthetic inductors}, i.e., circuits which artificially mimic the electrical behavior of conventional inductors. An example is the \emph{Antoniou circuit}~\cite{Antoniou_IEEE_1969}, which involves two operational amplifiers, a capacitor and four resistors, arranged as sketched in Fig.~\ref{fig:circ}a.
The equivalent inductance can be calculated as: 
\begin{equation}
L_{eq}=\frac{R_1\,R_3\,R_4\,C}{R_2}\,\,\,\,\,.
\label{eq:ant}
\end{equation}
Another major advantage of using this circuit is the fact that the value of the equivalent inductance can be modified by simply tuning one of the components. This feature is especially important in the context of our work, as tunability of the response is a key objective. In our case, as visible in the realization of the circuit on a breadboard shown in Fig.~\ref{fig:circ}b, the tunable component is the potentiometer corresponding to $R_1$, which can assume values in the $100$--$1000\,\mathrm{\Omega}$ range. The other circuital components are: $R_2=680\,\mathrm{\Omega}$, $R_3=1000\,\mathrm{\Omega}$, $R_4=1500\,\mathrm{\Omega}$, $C=0.22\,\mathrm{\mu F}$ (Mylar type), while, for the operational amplifiers, we select the OPA445 model (TI OPA445AP). Note that the amplifiers are powered by two DC power supplies (BK Precision 1667) arranged in series as to share the same ground and produce a voltage difference of $\pm30\,\mathrm{V}$. Note that the ground terminal coming from the power supplies is used as reference ground for the circuits as well as for the patches. More details on grounding are reported in the SD section. Given these components, the circuit can assume inductance values ranging from $0.0485\,\mathrm{H}$ to $0.485\,\mathrm{H}$ (from Eq.~\ref{eq:ant}). Since the average capacitance of the piezo patches is $C_{p,\,ave}= 2.137\,\mathrm{nF}$, the piezo+RL circuits can be tuned to resonate between $4950\,\mathrm{Hz}$ and $15600\,\mathrm{Hz}$ (from Eq.~\ref{eq:res}).

It is important to point out that the choice of circuit components is not accidental. In fact, in order to harness the full attenuation capability of the shunting strategy, it is fundamental to reduce as much as possible the \emph{parasitic} resistance of the circuit (i.e., the resistance that is inherent to the circuit components). The parasitic resistance can be measured experimentally as a function of the circuit components. In general, lower equivalent inductances lead to smaller values of the parasitic resistance; in light of this, working at high frequencies can be advantageous. For this reason, we decided to work in the $7$--$8.8\,\mathrm{kHz}$ frequency range. In any case, once the components are chosen and the circuits are assembled, it is suggested to test their electric response. Details on how to electrically test RL circuits are also discussed in the SD section.

\subsection{Shunting and ``screaming''}
\label{sec:scream}
While the beam comprises eighteen randomly-positioned patches, we only shunt seven of them (the results in this article refer to the activation of patches 1, 5, 8, 10, 12, 14 and 18, according to the numbering in Fig.~\ref{fig:beam}). Working with a subset of patches strengthens our claim regarding the non-influence of the patch positioning. It is worth pointing out that increasing the number of patches working at the same time is not a simple task. This is due to the fact that shunting multiple patches can lead to an unstable behavior of the system. This is especially observed when the series resistors $R_s$ (as previously mentioned, the circuit connected to each patch also includes a series resistor $R_s$) are zero or small. In an ideal scenario, it would be preferable not to use any series resistor, to enhance the attenuation behavior of each resonant circuit at its resonant frequency. However, without $R_s$, spurious vibrations are transmitted to the beam specimen and the patches can be heard ``screaming'', i.e., vibrating in the audible frequency range. Interestingly, this behavior is particularly strong when multiple shunts are tuned to resonate at the same frequency. This aspect is discussed more in detail in the SD section. In the following, to avoid instability phenomena at the frequencies of interest, we select $R_s=100\,\mathrm{\Omega}$.

\subsection{Signal generation and response acquisition}
\begin{figure} [!htb]
\centering
\includegraphics[scale=1.38]{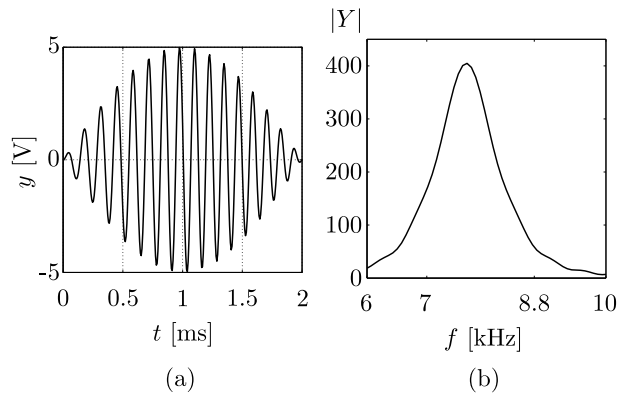}
\caption{Input signal: linear chirp with frequency content ranging from $7\,\mathrm{kHz}$ to $8.8\,\mathrm{kHz}$. (a) Time history. (b) Frequency spectrum.}
\label{fig:sig}
\end{figure}

To generate and acquire the signals, we resort to a data acquisition system (NI cDAQ-9174), equipped with an output module (NI 9263) and an input module (NI 9215), and connected to a laptop running the LabVIEW software. The input signal is generated in LabVIEW, sent through the output module to a power amplifier (Br\"uel \& Kj\ae r Type 2718) and then fed to a shaker (Br\"uel \& Kj\ae r Type 4810), which transmits it to the beam specimen through a stinger. To span the frequency range of interest (highlighted in Fig.~\ref{fig:beamtf}), we prescribe the signal shown in Fig.~\ref{fig:sig}a, i.e., a sine-modulated chirp with frequency content linearly increasing from $7\,\mathrm{kHz}$ to $8.8\,\mathrm{kHz}$ in $2\,\mathrm{ms}$. The frequency content of the signal is shown in Fig.~\ref{fig:sig}b.

\begin{figure*} [!htb]
\centering
\includegraphics[scale=1.38]{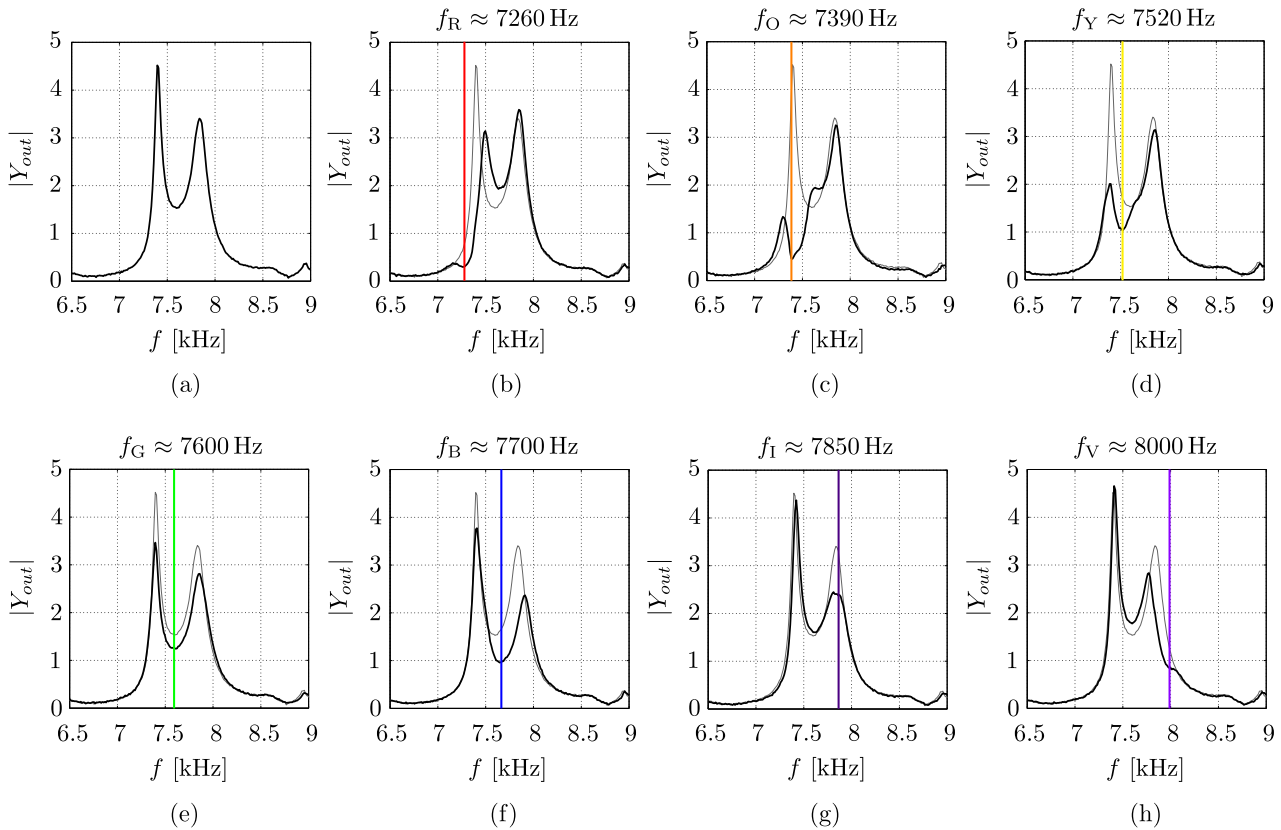}
\caption{Experimental results on single-frequency distillation; these spectra are obtained via Discrete Fourier Transform (DFT) of the transient signal acquired at the output accelerometer. (a) Short circuit. (b) RED shunt only. (c) ORANGE shunt only. (d) YELLOW shunt only. (e) GREEN shunt only. (f) BLUE shunt only. (g) INDIGO shunt only. (h) VIOLET shunt only. In (b-h), the vertical lines indicate the estimated tuning frequency of each shunt, while the thin gray profiles in the background correspond to the short circuit response, used as reference.}
\label{fig:fs}
\end{figure*}

To evaluate the system response, we analyze the data recorded by the output accelerometer, i.e., the one which is further away from the excitation source. The signal recorded by this accelerometer (PCB Model 352A73) is conditioned with a signal conditioner (PCB Model 482A22), sent to the input module and displayed in LabVIEW. The postprocessing of the data is performed in MATLAB. To reduce the noise, the signal corresponding to each separate experiment is acquired seven times and averaged.

\begin{figure*} [!htb]
\centering
\includegraphics[scale=1.38]{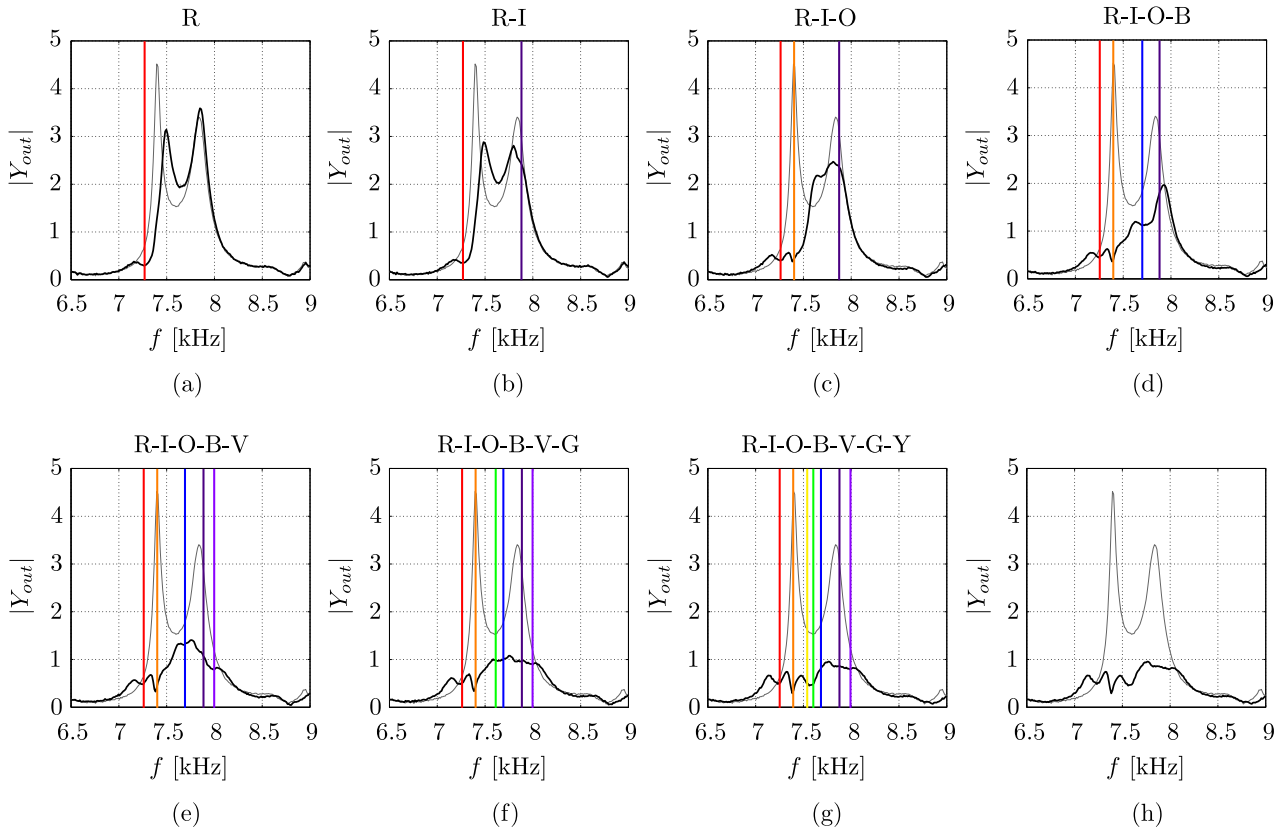}
\caption{Behavior of the rainbow trap in the frequency domain. The trap is activated following the sequence R-I-O-B-V-G-Y. (a-g) Results after each of the seven activation steps. The thin gray profiles in the background represent the frequency content of the signal recorded when the patches are all short circuited. (h) Final spectrum of the signal, obtained when all seven shunts are activated.}
\label{fig:f}
\end{figure*}

\section{Circuit tuning and single-frequency distillation}
\label{sec:res1}
Let us recall that our objective is to realize a tunable bandgap filter capable of operating over broadband spectra. This is the elasto-acoustic analog of a concept that, in optics and electromagnetism, has been originally termed ``rainbow trap''. The idea consists of shunting a discrete number of patches and tuning each one individually at different adjacent frequencies, such that the corrections are felt over slightly overlapping frequency bands. Here, we demonstrate the concept using seven of the amenable eighteen patches (1, 5, 8, 10, 12, 14 and 18). The individual tuning of the patches is performed by monitoring real-time changes in the frequency content of the signal recorded at the output accelerometer. Fig.~\ref{fig:fs}a represents the frequency content of the signal in the reference scenario, in which all patches are short circuited.
We can see that the spectrum of the applied chirp has morphed into the two-peak structure of Fig.~\ref{fig:fs}a, to reflect the presence of two resonances in the frequency response of the beam in the $7$--$8.8\,\mathrm{kHz}$ range (as shown in Fig.~\ref{fig:beamtf}). The thick black curves in Figs.~\ref{fig:fs}b-h represent the frequency content of the signals obtained by activating one patch at the time, superimposed on top of the response of the short-circuited beam (thin gray line). We can see that each shunt has a very localized behavior, being able to \emph{distill} (subtract) the energy associated with a very narrow frequency band. In analogy with the spectrum of visible light rainbows, and in order to maintain a convenient and visually friendly taxonomy of the frequency sequence, we name each of the seven shunts as one of the seven colors of a discrete (Newtonian) rainbow: the RED (R) shunt resonates at $f_{\mathrm{R}}\approx 7260\,\mathrm{Hz}$ (Fig.~\ref{fig:fs}b), the ORANGE (O) at $f_{\mathrm{O}}\approx 7390\,\mathrm{Hz}$ (Fig.~\ref{fig:fs}c), the YELLOW (Y) at $f_{\mathrm{Y}}\approx 7520\,\mathrm{Hz}$ (Fig.~\ref{fig:fs}d), the GREEN (G) at $f_{\mathrm{G}}\approx 7600\,\mathrm{Hz}$ (Fig.~\ref{fig:fs}e), the BLUE (B) at $f_{\mathrm{B}}\approx 7700\,\mathrm{Hz}$ (Fig.~\ref{fig:fs}f), the INDIGO (I) at $f_{\mathrm{I}}\approx 7850\,\mathrm{Hz}$ (Fig.~\ref{fig:fs}g) and the VIOLET (V) at $f_{\mathrm{V}}\approx 8000\,\mathrm{Hz}$ (Fig.~\ref{fig:fs}h). These resonant frequencies are estimated by inspecting Figs.~\ref{fig:fs}b-h and are highlighted by colored vertical lines. Also note that the R shunt corresponds to patch 14, the O to 8, the Y to 18, the G to 12, the B to 1, the I to 10 and the V to 5. By comparing the results produced by individual shunts, we can see that some of them cause stronger attenuation than others. In our opinion, this is not influenced by the position of the patch and it might be predominantly due to bonding imperfections (which could reduce the effective contact area of some of the patches).

\section{The rainbow trap}
\label{sec:res2}
In this section, we illustrate the result of the simultaneous activation of multiple patches and we provide a demonstration of our tunable electromechanical rainbow trap, analyzing its effects both on the frequency content of the signal and on the shape of the wavepacket.

\subsection{Frequency response - Bandgap effect}
\label{sec:f}
To activate the broadband trap, we short circuit all the patches, we turn on the DC power supplies and then we connect all the shunts one after the other, according to a pre-determined \emph{activation sequence}. Note that any activation sequence is bound to produce the same net result, due to the linearity of our system; to prove this point, in the SD section, we report the results obtained with a different sequence, which indeed yields the same overall signal attenuation. For completeness, we report the data recorded at each activation step. The frequency content of the signal recorded at the output accelerometer, for a R-I-O-B-V-G-Y activation sequence, is reported in Fig.~\ref{fig:f}. In Figs.~\ref{fig:f}a-g, the colored vertical lines indicate the frequencies at which the shunts activated at each step of the sequence are tuned (the lines correspond to the tuning frequencies estimated in Fig.~\ref{fig:fs}). As we go through the shunting sequence, the energy associated with the wave packet is gradually distilled: each shunt absorbs the energy corresponding to frequencies close to its resonance, until we reach the final attenuation stage shown in Fig.~\ref{fig:f}h. It is interesting to notice how the activation of some patches causes the frequency content of the signal to be stretched, compressed and deformed, apparently deploying some of the energy from the corrected region to a neighboring frequency interval. For example, this is visible in Fig.~\ref{fig:f}a, where the RED circuit causes frequencies around $7.7\,\mathrm{kHz}$ to become more favorable than in the reference case. Fig.~\ref{fig:f}h allows to fully appreciate the broadband wave attenuation capability of our rainbow trap; it also confirms how the periodicity of the patch placement and the homogeneity of the shunts characteristics are not required in a transient wave propagation problem involving RL shunts. This is due to the fact that the traveling packet, which has local support, propagates along the beam and is therefore forced to interact sequentially with all the patches, thus feeling the effect of each correction. Moreover, due to its short wavelength characteristics, the wave has the opportunity to impart significant deformation to each patch. In contrast, in vibration attenuation problems (steady-state conditions), the possibility for the patches to undergo significant levels of strain at the frequencies corresponding to the peaks that we intend to suppress is directly related to the modal characteristics of the structure. In such scenario, the arrangement of patches would need to be carefully designed. 

\begin{figure*} [!htb]
\centering
\includegraphics[scale=1.38]{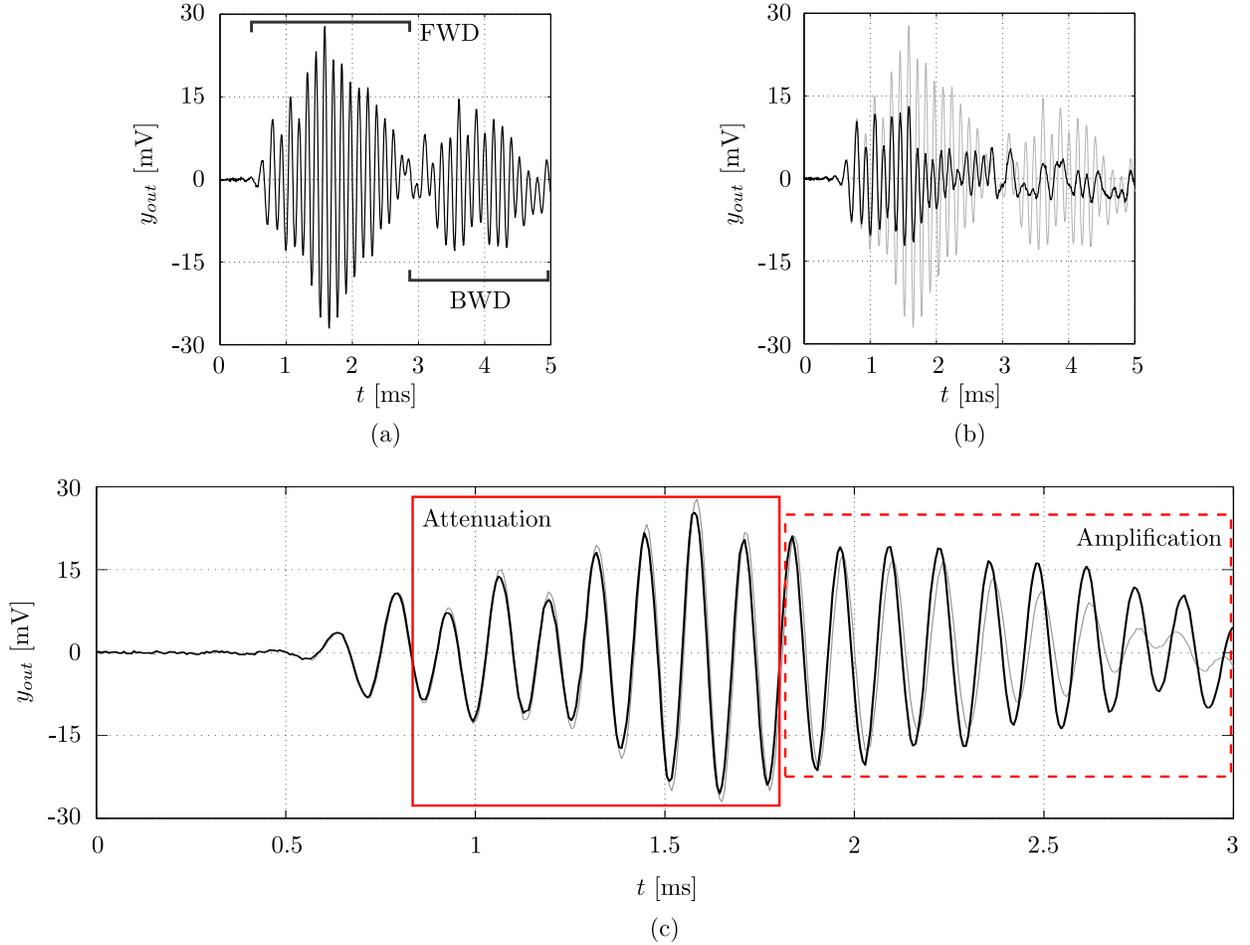}
\caption{Effect of the rainbow trap on the shape of the wavepacket. (a) Short circuit scenario---the backward (BWD) part is due to boundary reflection. (b) Signal comparison between fully-activated rainbow trap (black line) and short circuit case (gray line), highlighting broadband attenuation and distortion. (c) Detail of the effect of the RED shunt (black line), compared to the short circuit case (gray line), highlighting the coexistence of zones of attenuation and zones of amplification.}
\label{fig:t}
\end{figure*}

\subsection{Time domain considerations - Packet distortion}
\label{sec:t}
Most of the existing literature on RL shunts focuses on spectral effects and frequency manipulation. In this work, we attempt to provide an additional angle to the problem, by monitoring how resonant shunts can affect the morphological characteristics of a traveling signal. Since our working signal is a chirp, with a frequency content that evolves across the packet, it is interesting to explore how the effects of specific frequency distillations are recorded in different (predictable) zones of the signal (e.g. front or tail) and how they manifest as localized modifications of the packet envelope. In Fig.~\ref{fig:t}a, we report the wavepacket recorded when all the patches are short circuited. In the $0$--$3\,\mathrm{ms}$ interval, we can identify a first packet, corresponding to the signal traveling away from the source, here termed forward (FWD) packet. The oscillations starting approximately at $3\,\mathrm{ms}$, on the other hand, correspond to the signal reflected by the clamped end of the beam and traveling back towards the source, here termed backward (BWD) packet. We can see that, even without shunting, the FWD packet already features some distortion with respect to the prescribed signal (shown in Fig.~\ref{fig:sig}a), due to the inherent dispersive characteristics of the beam and its bimodal response in the frequency region of interest. In Fig.~\ref{fig:t}b, we compare the signal recorded when the rainbow trap is completely activated (black line) against the short circuited case (gray line). This representation confirms how the shunting strategy causes a generalized de-energization of the signal and indicates that the rainbow trap effectively acts as an \emph{acousto-elastic signal jammer}.

It is especially interesting to analyze the effects of each individual step of the activation sequence on the wavepacket distortion. For the sake of brevity, we only report the analysis of the first step of the sequence, corresponding to the application of the RED shunt (Fig.~\ref{fig:t}c); the black line corresponds to the wavepacket upon shunting and the gray line reflects the short circuit case, which is taken as reference to quantify the influence of the frequency distillation. In the interest of clarity, we limit the inspection to the FWD portion of the signal, although similar considerations can be made for the reflected wave. In the time interval of interest, we can recognize two distinct zones with modified features. Between $0.8\,\mathrm{ms}$ and $1.8\,\mathrm{ms}$ the wave is attenuated, as highlighted by the amplitude reduction of the wave crests. This conforms to the fact that the frequency content of this portion of the packet lives in the neighborhood of the resonance frequency of the RED shunt. In contrast, an increase in the crests amplitude indicates that the wave is amplified in the $1.8$--$3\,\mathrm{ms}$ interval. This observation is consistent with the global picture provided by the spectrum of Fig.~\ref{fig:f}a: while the signal is dynamically damped in the neighborhood of the natural frequency of the resonant shunt, it is locally amplified at frequencies above resonance, which are here located towards the tail of the packet. A complete description of the effects of the other steps in the activation sequence is reported in the SD section. In summary, due to the activation of each shunt, we observe a localization of the amplitude correction effects in the signal which is consistent with the position of the frequencies in the chirp time history. This could open avenues towards the design of structures featuring RL shunts that could be programmed to impart desired morphological characteristics and engineered distortion patterns to the envelope of broadband signals.

\section{Conclusions}
\label{sec:con}
To summarize, in this work we have demonstrated the wave attenuation capabilities of a random array of piezo elements bonded to a waveguide (beam) and shunted with RL circuits. By letting the shunts resonate at adjacent frequencies, we obtained broadband attenuation of a traveling wavepacket. We have revised the concept of ``organized disorder'' in the context of shunted piezoelectrics, thus adding a layer to the outstanding versatility of shunted piezo elements and their applicability in the design of tunable metamaterials.

\ack
We acknowledge the support of the National Science Foundation (grant CMMI-1266089). Davide Cardella also acknowledges the support of the Outgoing office of Politecnico di Torino (through the Master Thesis Abroad Scholarship). We are indebted to Lauren E. Linderman and Paul Bergson (University of Minnesota) for their support with the experimental equipment. We are grateful to Iacopo Gentilini and Douglas Isenberg (Embry-Riddle Aeronautical University, Prescott) for their helpful insight and assistance with circuitry during the initial stages of our work. Finally, we thank Benjamin S. Beck (Pennsylvania State University) for an enlightening discussion on shunted piezoelectrics.

\clearpage
\section*{References}
\bibliographystyle{iopart-num}
\balance
\providecommand{\newblock}{}


\clearpage
\nobalance
\section*{\Large Supplementary Data (SD)}
\renewcommand{\thefigure}{S\arabic{figure}}
\renewcommand{\theequation}{S\arabic{equation}}
\renewcommand{\thepage}{S\arabic{page}}
\renewcommand{\thesection}{S\arabic{section}}
\setcounter{figure}{0}
\setcounter{page}{1}
\setcounter{section}{0}

\section{Bonding the patches}
\begin{figure} [!htb]
\centering
\includegraphics[scale=1.38]{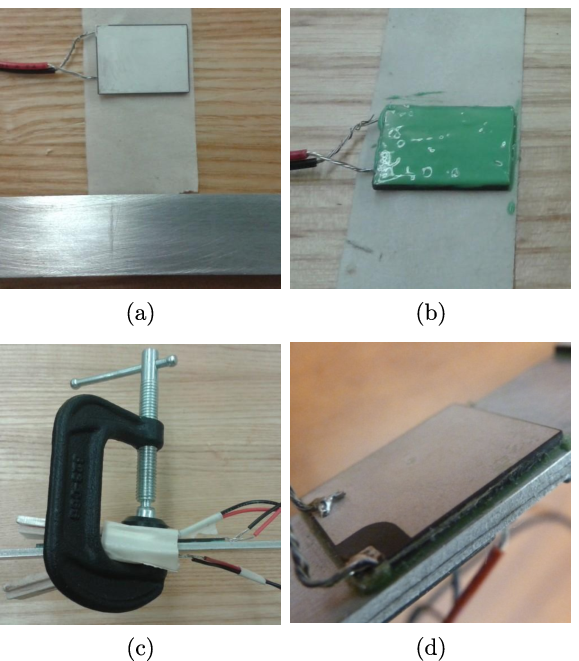}
\caption{Stages of the bonding procedure. (a) Surface cleaning. (b) Glue application. (c) Pressure application. (d) Final result showing the soldered wires.}
\label{fig:bond}
\end{figure}

A proper bonding of the piezoelectric patches to the substrate is a key aspect for the realization of successful experiments involving shunting circuits [Ducarne, \emph{Mod\'elisation et optimisation de dispositifs non-lin\'eaires d’amortissement de structures par syst\'emes pi\'ezo\'electriques commut\'es}, PhD Thesis, CNAM, 2009]. As piezoelectric materials react to strain by creating a voltage, a strong coupling between patch and substrate ensures that the generated voltage is large enough to trigger the desired effects in the shunt circuits. For this reason, we report a few useful tips on how to properly bond PZT patches to an Aluminum substrate. In our work, we use a two-part epoxy glue (3M Scotch-Weld 1838 B/A) as bonding agent. The bonding procedure consists of three basic stages: \emph{surface cleaning}, \emph{glue application} and \emph{pressure application}. These stages are shown in Fig.~\ref{fig:bond}. Note that it is suggested to solder the wires to the electrodes of the patch only after the bonding procedure.

\subsection{Surface cleaning}
A proper cleaning of the surfaces to be bonded (both the patch's beam-side electrode and the beam substrate) is crucial to eliminate impurities, which might compromise the quality of the bonding. As far as the piezo patch is concerned, we suggest to apply tape on the top-side electrode, as shown in Fig.~\ref{fig:bond}a; this should facilitate the patch positioning, without any need to touch the treated surface. To clean the electrode and to dissolve traces of grease, we used an acetone solution. Then, the patch must be washed with water. It is very important not to use sandpaper on the patch; this could result in permanent damage to the thin electrodes. To prepare the surface of the beam, we follow a different procedure. First, we use fine-grade sandpaper to create micro-asperities in the contact area: this will increase the effectiveness of the bonding. This etching process can be enhanced by also applying a diluted solution of sulfuric acid, which should then be removed with an acid removal solution and water. From this point on, touching the contact area should be avoided. The steps that follow should be performed quickly to minimize the amount of aluminum oxide formed on the beam surface.

\subsection{Epoxy glue application}
At this stage, we can prepare the mixture of epoxy glue and reagent (for the 3M Scotch-Weld 1838 B/A Epoxy, it is suggested to use equivalent amounts in weight). After the mixture is created, we apply a thin layer of it to the beam-side electrode of the patch, as shown in Fig.~\ref{fig:bond}b. This application step can be performed with a wooden stick or a similar tool. It is important to try to spread the layer along a single direction. Note that this step is very crucial: the quality of the bonding will strongly depend on whether the glue has been applied evenly or not. 

\subsection{Pressure application}
At this stage, we can use the previously-applied tape segment to lift the patch from the table and to place it in the desired position on the substrate. Now we can apply pressure, using a weight or a clamp (as shown in Fig.~\ref{fig:bond}c), to reduce the thickness of the glue layer and to guarantee that bonding will take place. This pressure needs to be applied for the whole ``resting time'' indicated by the epoxy manufacturer. After a couple of hours, when the glue is not yet solid, we suggest to remove the clamp and check the patch positioning; at this stage, small displacements are still allowed. Moreover, this is a good time to remove excess epoxy glue. Note that, if using a conductive glue, it is fundamental to avoid glue on the top-side electrode; this would cause an unwanted short-circuiting of the patch. 

\section{Testing the circuits}
\begin{figure} [!htb]
\centering
\includegraphics[scale=1.38]{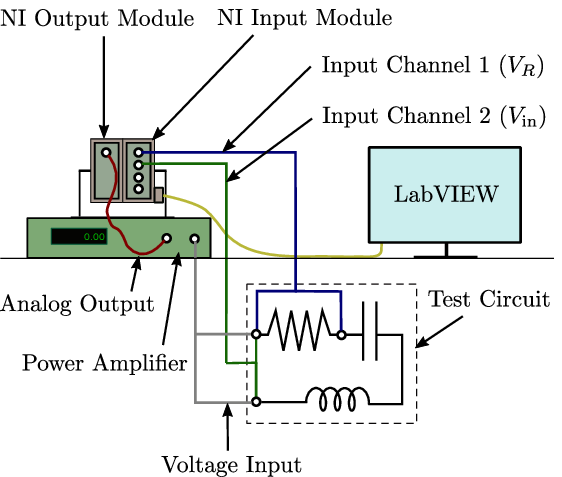}
\caption{Sketch of the experimental setup used to test the Antoniou circuits.}
\label{fig:circDAQ}
\end{figure}

\begin{figure} [!htb]
\centering
\includegraphics[scale=1.38]{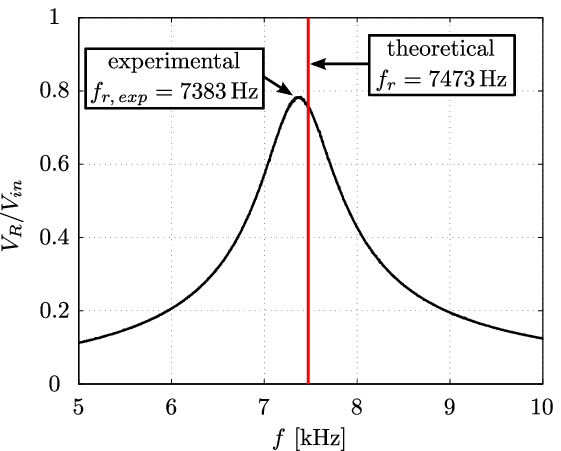}
\caption{Testing the Antoniou circuit of the RED shunt. The black line represents the experimental frequency response function of the test resonant circuit. The red line indicates the resonance predicted by Eq.~\ref{eq:rres}.}
\label{fig:circR}
\end{figure}

As mentioned in the main article (Sec.~\ref{sec:circ}), we suggest to test the Antoniou circuits electrically before attempting to connect them to the patches; this strategy allows to single out issues related to the circuitry. The experimental setup we used to perform these tests is shown in Fig.~\ref{fig:circDAQ}.
The goal is to test the functioning of the Antoniou circuit, whose inductance is labeled $L_{eq}$. To do so, we connect it to a test resistance $R_t$ and to a test capacitance $C_t$ [Viana and Steffen, \emph{J. Braz. Soc. Mech. Sci. Eng.} 28, 293--310, 2006]; the $R_t$$L_{eq}$$C_t$ circuit resonates at:
\begin{equation}
f_{r}=\frac{1}{2\,\pi}\sqrt{\frac{1}{L_{eq}\,C_t}}\,\,\,\,\,.
\label{eq:rres}
\end{equation}
The circuit is tested by subjecting it to a multitone input voltage containing a superposition of harmonics between $5\,\mathrm{kHz}$ and $10\,\mathrm{kHz}$ with a frequency increment of $0.1\,\mathrm{Hz}$ (generated in LabVIEW, sent through the output module of the DAQ to the power amplifier and then fed to the resonant circuit). A frequency response function of the circuit is then calculated by dividing the voltage measured at $R_t$ ($V_R$) by the input voltage ($V_{in}$). If the synthetic inductor works properly, the response is characterized by a single peak corresponding to the resonance of the circuit. The response of the Antoniou circuit in one of the shunts used in this work (the RED shunt), coupled to a test resistance $R_t=1\,\mathrm{k\Omega}$ and to a test capacitance $C_t=2.2\,\mathrm{nF}$ is shown in Fig.~\ref{fig:circR}.
Note that, to calculate the theoretical $f_r$, we used measured values of $L_{eq}$ ($0.2189\,\mathrm{H}$, obtained from Eq.~\ref{eq:ant} by setting $R_1=440\,\mathrm{\Omega}$, $R_2=666\,\mathrm{\Omega}$, $R_3=981\,\mathrm{\Omega}$, $R_4=1480\,\mathrm{\Omega}$ and $C=228.2\,\mathrm{nF}$) and $C_t$ ($2.072\,\mathrm{nF}$), which slightly differ from the nominal ones. We can see that the experimental result in terms of resonant frequency agrees well with the theoretical one. Differences in measured and calculated $f_r$ are inevitable, and could be due to the presence of additional (parasitic) resistances in the Antoniou circuit.

\section{Minimizing the parasitic resistance}
The \emph{parasitic} resistance $R_p$ is inherent to the circuit components. In a synthetic inductor, $R_p$ is due to the presence of capacitors and operational amplifiers. $R_p$ is usually considered responsible for the differences between the nominal and the actual behavior of the Antoniou circuits. For example, this is the case for the results shown in Fig.S3: the parasitic resistance is responsible for the discrepancy in resonance frequency, and for the fact that the maximum amplitude of the FRF is not equal to 1, as predicted by the analytical RLC circuit model. In the literature the parasitic resistance has been quantified by modeling it as a resistance added in series (or in parallel) to the equivalent inductor [Viana and Steffen, \emph{J. Braz. Soc. Mech. Sci. Eng.} 28, 293--310, 2006]. These simple models provide a tractable avenue to capture a few important effects of $R_p$ on the circuit behavior. For example, the series model explains the amplitude mismatch in Fig. S3, where the peak of the FRF is below the nominal value 1 predicted by the theoretical circuit model; based on this argument, one can obtain an easy calibration of the parameter $R_p$ targeting the specific amplitude. On the other hand, it fails at capturing other features in the response, such as the shift in frequency. The frequency shift can be qualitatively explained by recognizing that the parasitic resistance arises internally to the Antoniou circuit and is, in some sense, intertwined to its inherent components, in a way that would be only partially captured by modeling the effect by an external impedance (be it series or parallel). A parasitic resistance in the components would affect the expression of the equivalent capacitance in the Antoniou circuit formulation, which, in turn, would surely result in a correction of the resonant frequency. This description, however, would necessarily involve a much less tractable circuital model (in which multiple internal sources of parasitic resistance would have to be considered and weighted appropriately) and would not be easily amenable to model calibration.

In general, it is best to design the circuits as to reduce as much as possible $R_p$. In fact, as $R_p$ gets larger, we might be facing a scenario in which the Antoniou circuits stop working properly; this could cause the resonant shunt to lose its resonant behavior. Here are some practical considerations which we followed to minimize the parasitic resistance: first, it is best to place a capacitor with low value of $C$ in the Antoniou circuit; second, the equivalent inductance of the Antoniou circuit $L_{eq}$ should be as small as possible; third, we suggest to use Mylar capacitors, which seem to be characterized by smaller inherent resistances with respect to the electrolytic ones. In a shunting scenario, the second consideration implies that working at higher frequencies is more convenient (in fact, for a fixed capacitance of the piezo $C_p$, higher frequencies imply smaller values of $L_{eq}$).

\section{Grounding}
\begin{figure} [!htb]
\centering
\includegraphics[scale=1.38]{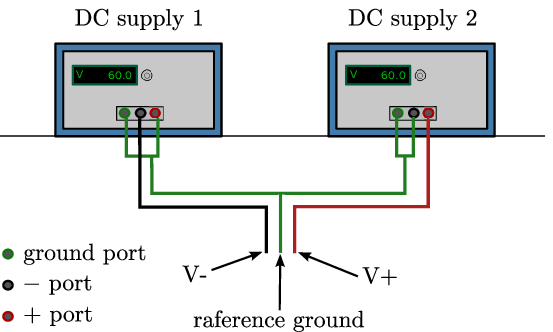}
\caption{Schematic of the DC power supplies, showing how to obtain a reference ground.}
\label{fig:DC}
\end{figure}

While a physical inductor has a floating reference point, the synthetic inductor needs a fixed reference ground. The ground of the circuit needs to be the same as the ground of the DC power supply used to power the operational amplifiers. In our case, we use two DC power supplies connected as shown in Fig.~\ref{fig:DC}.
This series connection allows to generate a $\pm 60\,\mathrm{V}$ voltage (in our experiments, we use $\pm 30\,\mathrm{V}$ only), while also providing a reference ground pin. This ground pin is used to ground the synthetic inductors as well.

When using the Antoniou circuits in the resonant shunts, it is also important to have them share the same ground as the patches. For this reason, the beam-side electrode of each patch is connected to the reference ground. Since the epoxy used for bonding does not display excellent conductivity, this represents the best strategy to assure that the patches share the same ground. If conductive epoxy were to be used, it would have been sufficient to connect the beam to the reference ground.

\begin{figure} [!htb]
\centering
\includegraphics[scale=1.38]{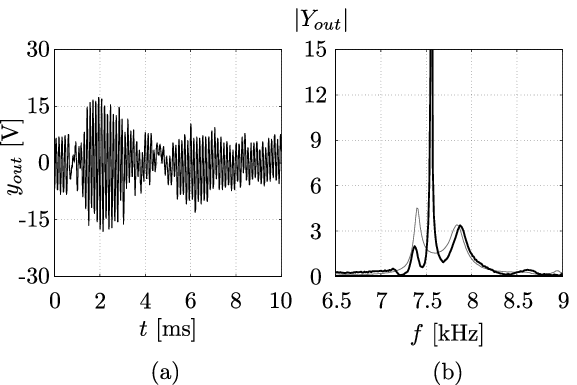}
\caption{Response, recorded at the output accelerometer, when circuits 12, 14 and 18 are simultaneously shunted, without series resistors. (a) Time response. (b) Frequency spectrum.}
\label{fig:noisy}
\end{figure}
\begin{figure*} [!htb]
\centering
\includegraphics[scale=1.38]{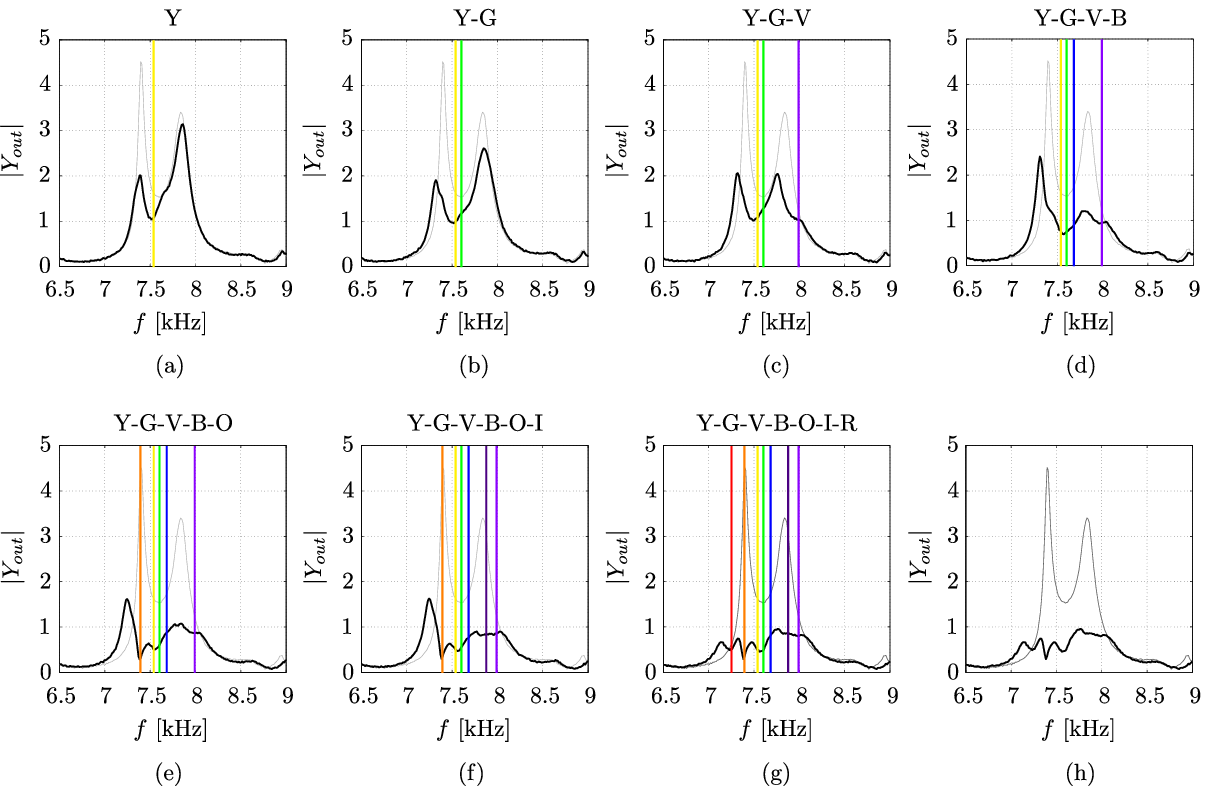}
\caption{Behavior of the rainbow trap in the frequency domain (to be compared with the alternate sequence shown in Fig.~7 of Sec.~4.1). The trap is activated following the sequence Y-G-V-B-O-I-R. (a-g) Results after each of the seven activation steps. The thin gray profiles in the background represent the frequency content of the signal recorded when the patches are all short circuited. (h) Final spectrum of the signal, obtained when all seven shunts are activated.}
\label{fig:f2}
\end{figure*}

\section{Instabilities}
In Sec.~\ref{sec:scream}, we discussed how a series resistor $R_s$ is always needed when using resonant shunts, to suppress instability phenomena that take place when multiple patches are shunted. These instabilities manifest in several ways. The most characteristic indicator is a ``screaming'' sound that can be heard coming from some of the shunted patches; this noise is caused by strong vibration of the patches in the audible frequency range. Another manifestation of this unstable behavior is the heating up of the circuit components, which is to be avoided at all costs, as it can quickly (and irreversibly) damage the components. To investigate this phenomenon, we recorded the signal at the output accelerometer when patches 12, 14 and 18 were shunted at the same time, without any resistor $R_s$ in the shunts. Note that, due to the fact that circuits were heating up during the signal acquisition, we did not perform multiple measurements for this experiment. The results in terms of time signal and frequency spectrum are shown in Fig.~\ref{fig:noisy}. We can see that the time response is characterized by unexpected oscillations starting right at the beginning of the acquisition. This translates into a very strong resonance peak recorded around $7550\,\mathrm{Hz}$. Note that this peak reaches amplitudes far higher than those recorded in the short circuit case (the short circuit spectrum is also shown in Fig.~\ref{fig:noisy}(b), as a thin gray line). Not surprisingly, the frequency at which the unexpected resonance occurs is very close to the resonant frequency of one of the connected circuits (circuit 18, which resonates at $f_Y\approx 7520\,\mathrm{Hz}$). While we are not sure about the reason behind this unstable behavior, placing a large-enough series resistance $R_s$ in the shunts appears to be an effective solution to the problem. Note that this unstable behavior is essentially not discussed in the literature, except for a few useful comments in the work of dell'Isola et al. [dell'Isola et al., \emph{Smart Mater. Struct.} 13, 299, 2004].

\section{Activation sequence}
As briefly stated in Sec.~\ref{sec:f}, the results we obtain through the activation of the rainbow trap do not depend on the activation sequence. This is a reflection of the linearity of our system, guaranteed by working with small strains and low voltages. To substantiate this claim, in Fig.~\ref{fig:f2} we report the results (in terms of spectral manipulation of the wavepacket) obtained following an alternate activation sequence from the one illustrated in the main text (here we use Y-G-V-B-O-I-R instead of R-I-O-B-V-G-Y). 
It can be seen that the final result in Fig.~\ref{fig:f2}h is indeed identical to that of Fig.~\ref{fig:f}h.

\section{Wave packet manipulation}
\begin{figure*} [!htb]
\centering
\includegraphics[scale=1.38]{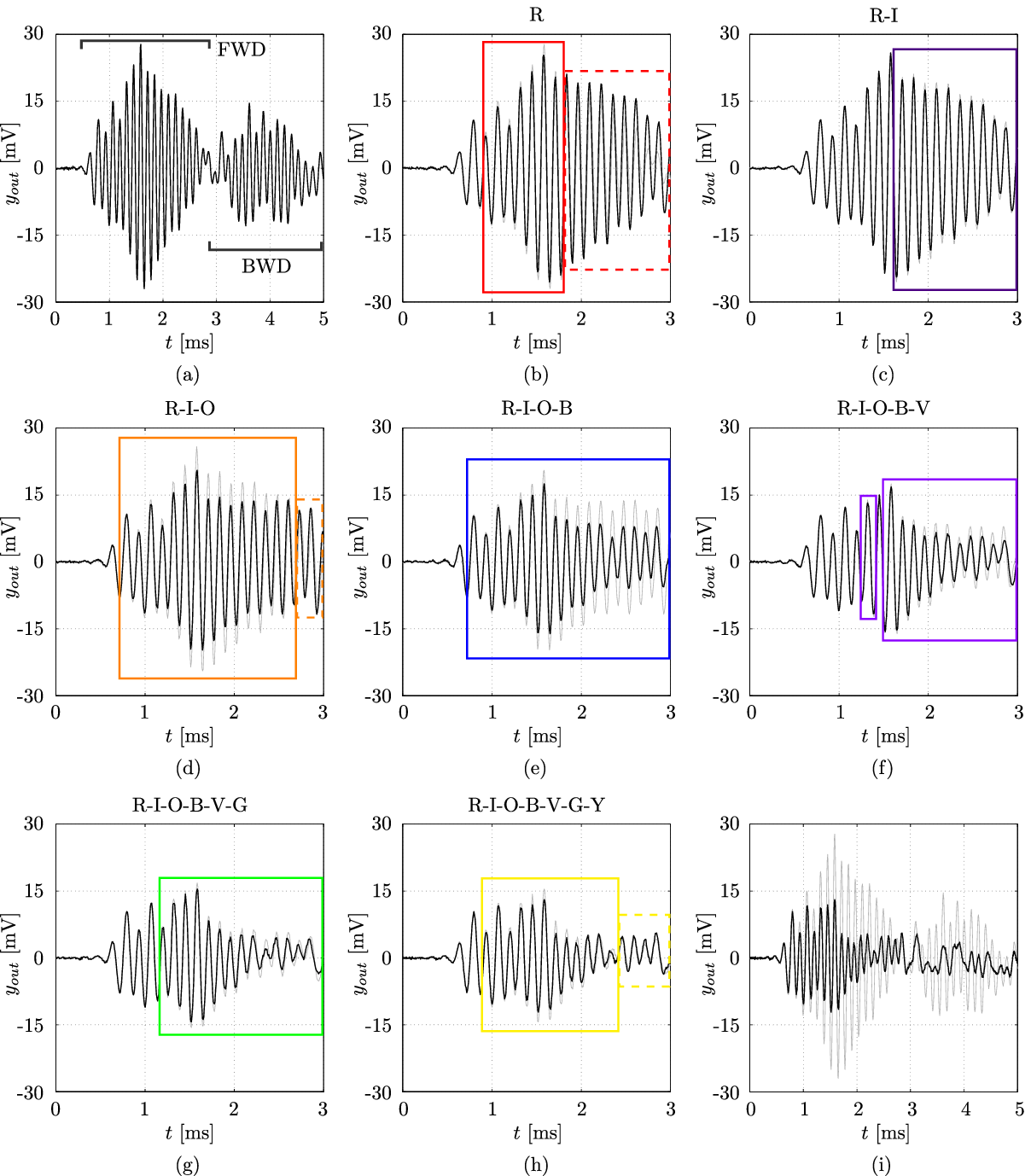}
\caption{Effect of the rainbow trap on the shape of the wavepacket. (a) Short circuit scenario. (b-h) Wavepacket after each of the seven activation steps (black lines). To provide a visual comparison, the gray lines in (b-h) represent the packet profile at the previous step; the boxed regions highlight, at each step, individual distorted features (solid line: attenuation, dashed line: amplification). (i) Comparison between the signal obtained when the trap is fully activated (black line) and the signal in the short circuit case (gray line), highlighting the global distortion due to the rainbow trap.}
\label{fig:t2}
\end{figure*}

In Sec.~\ref{sec:t} we discussed the influence on the wavepacket of the shunting circuits, using the RED shunt as an example. For the sake of completeness, we here report the intermediate effects of the other steps in the R-I-O-B-V-G-Y sequence. The time signals (FWD portion only) recorded at each step of the activation sequence are shown in Figs.~\ref{fig:t2}b-g. In each of these plots, the signal recorded for the current activation step (black line) is compared to the signal recorded at the previous step (gray line). The regions of the signal undergoing modifications are highlighted by colored boxes (boxes with solid contours correspond to attenuation regions, while dashed contours correspond to amplification regions).

\end{document}